%% file: main.tex
\documentclass[10pt, a4paper, twocolumn, showabstract]{naverlabseurope}

\usepackage{lipsum}
\usepackage{multicol}
\usepackage{tikz,tkz-kiviat,pgfplots}
\usepackage{tabularx}
\usepackage{booktabs}
\usepackage{amssymb}
\usepackage{amsmath}
\usepackage{float}
\usepackage[utf8]{inputenc}
\usepackage{microtype}
\usepackage{graphicx}
\usepackage{hyperref}
\usepackage{natbib}
\DeclareUnicodeCharacter{2588}{\rule{1ex}{1.2ex}}

\title{Naver Labs Europe @ WSDM CUP | Multilingual Retrieval}

\correspondingauthor{thibault.formal@gmail.com | $^{\dagger}${\bf Work done while at Naver Labs Europe}.}

\authors{Thibault Formal$^{\dagger}$ \authsep Maxime Louis \authsep \authsep Hervé Déjean \authsep Stéphane Clinchant}
\affiliations{NAVER LABS Europe}
\website{}
\websiteref{}

\begin{abstract}

This report presents our participation to the WSDM Cup 2026 shared task on multilingual document retrieval from English queries. The task provides a challenging benchmark for cross-lingual generalization. It also provides a natural testbed for evaluating SPLARE, our recently proposed learned sparse retrieval model, which produces generalizable sparse latent representations and is particularly well suited to multilingual retrieval settings.

We evaluate five progressively enhanced runs, starting from a SPLARE-7B model and incorporating lightweight improvements, including reranking with Qwen3-Reranker-4B and simple score fusion strategies. Our results demonstrate the strength of SPLARE compared to state-of-the-art dense baselines such as Qwen3-8B-Embed. More broadly, our submission highlights the continued relevance and competitiveness of learned sparse retrieval models beyond English-centric scenarios.

\end{abstract}

\begin{document}

\maketitle

\input{sections/intro}
\input{sections/experiments}
\input{sections/conclusion}

\clearpage
{
    \small
    \bibliographystyle{ieeenat_fullname}
    \bibliography{bib}
}

\clearpage
\appendix

\input{tables/examples-2}

\end{document}

%% file: sections/intro.tex
\section{Setting}\label{sec:setting}

\paragraph{\bf Task Description.} We describe our approach to the WSDM Cup 2026 shared task on multilingual document retrieval. The task requires participants to develop systems that receive English queries and retrieve relevant documents from a multilingual collection of approximately 10 million documents, comprising Chinese (3.1M), Persian (2.2M), and Russian (4.6M) texts. Systems are evaluated using nDCG@20.

\paragraph{Learned Sparse Retrieval.}

Recently, Learned Sparse Retrieval (LSR) models such as SPLADE~\citep{10.1145/3477495.3531857,10.1145/3634912} have struggled to remain competitive on emerging large-scale benchmarks like Multilingual MTEB~\citep{enevoldsen2025mmteb}. In contrast, dense embedding models—particularly those built upon Large Language Models—consistently achieve strong performance across these evaluations~\citep{lee2025geminiembeddinggeneralizableembeddings,qwen3embedding}. As a result, much of the research community’s attention has shifted toward dense approaches. Beyond their empirical effectiveness, dense models offer a conceptually simple and unified framework: inputs are projected into a shared latent space, enabling straightforward optimization with contrastive objectives and seamless extension to multilingual and cross-lingual settings. Their ease of training, scalability with large pre-trained backbones, and good generalization properties have made them the default choice for modern retrieval systems. 

We hypothesize that this gap may be partly attributable to structural limitations of LSR approaches, most notably their reliance on the fixed vocabulary inherited from the backbone language model to construct representations. This constraint makes adaptation and training for multilingual or cross-lingual settings more challenging~\citep{10.1145/3539618.3591644}. In recent work, MILCO~\citep{nguyen2026milco} introduced an alternative strategy based on English-centric vocabulary mapping of learned representations, combined with alignment pretraining to mitigate cross-lingual discrepancies. This approach has demonstrated promising results on datasets like MIRACL~\citep{zhang-etal-2023-miracl} relative to comparable dense embedding methods and represents a competitive solution for multilingual LSR~\citep{lawrie2025neuclirbenchmodernevaluationcollection}.

\paragraph{\bf SPLARE.}

Concurrently with MILCO, we have recently introduced SPLARE~\citep{formal2026learning}, a new  learned sparse retrieval approach that uses sparse autoencoders to map inputs into a unified, language-agnostic space of sparse latent features. We present examples on development queries in Tables~\ref{tab:table-example} and~\ref{tab:table-example-2}, illustrating the shared latent space of concepts in which the model operates, independent of language.
SPLARE is trained on a large-scale open-source multilingual corpus comprising several English datasets (e.g., MS MARCO and NQ), an extensive collection of Chinese datasets, as well as the training splits of MIRACL and Mr. TyDi~\citep{zhang-etal-2021-mr}\footnote{The dataset is publicly available on HuggingFace at \texttt{hanhainebula/bge-multilingual-gemma2-data}.}.
Compared to equivalent SPLADE models operating over a fixed LLM vocabulary, SPLARE demonstrates consistently stronger performance, particularly on multilingual retrieval benchmarks such as MIRACL  and cross-lingual tasks such as XTREME-UP~\citep{ruder-etal-2023-xtreme}. On the latter, its performance approaches that of Gemini-based dense embeddings (see~\citet{formal2026learning} for detailed comparisons and analyses). These results motivated our participation to the WSDM Cup.

\input{tables/examples-1}

\paragraph{Our participation.}

Our primary objective in participating is to assess the effectiveness of SPLARE in a novel multilingual setting. To this end, we evaluate five progressively enhanced configurations, starting from a baseline SPLARE-7B retrieval model and incrementally incorporating lightweight improvements, such as reranking. We do not use development labels for fine-tuning and rely largely on default hyperparameters for all experiments, including indexing with Seismic~\citep{bruch2024seismic} and score fusion. We conduct pure multilingual retrieval and reranking, without leveraging the English-translated version of the collection.

We concatenate document titles with their corresponding contents and apply SPLARE (and SPLADE) out of the box with a maximum input length of 1024 tokens. For reranking, we extend the context window to 2048. For LSR models (both SPLARE and SPLADE), we submit runs based on non-pooled representations to maximize retrieval effectiveness (see Sections~\ref{sec:run1} and~\ref{sec:2.1}). Efficiency considerations fall outside the scope of the challenge objectives. Nevertheless, system efficiency could be improved through techniques such as Top-K pooling (see, e.g., Table~\ref{tab:table-splare} and the extended discussion in~\citet{formal2026learning}) or methods such as two-step SPLADE~\citep{10.1007/978-3-031-56060-6_23}. These approaches typically involve a modest trade-off in effectiveness.

We report nDCG@20 on the development set for each of the five settings, and we provide in Table~\ref{tab:table-comparison} a comparison to strong baselines evaluated in~\citet{lawrie2025neuclirbenchmodernevaluationcollection}. {\bf Overall, SPLARE achieves substantial improvements (+7.5 nDCG@20 over Qwen3-8B-Embed) and surpasses the performance of a full system incorporating reranking. Further gains obtained through rank fusion and reranking yield state-of-the-art results.}

\input{tables/comparison}

%% file: tables/examples-1.tex
\begin{table*}[t]
\centering
\begin{tabularx}{0.85\linewidth}{l}
\toprule
{\small $\mathcal{Q} \rightarrow$ Iranian female athletes refugees I am looking for stories about Iranian female athletes} \\
{\small  who seek asylum in other countries.} \\
\\
\texttt{\footnotesize references to athletes and athleticism                       | ████████████████████ 3.69} \\
\texttt{\footnotesize references to asylum seekers and related terms               | ███████████████████ 3.67} 
 \\
\texttt{\footnotesize various forms of the word "story" and its related concepts   | ███████████████████ 3.66} \\
\texttt{\footnotesize references to specific locations and their geopolitical cont | ███████████████████ 3.64} \\
\texttt{\footnotesize references to Iran and related entities or subjects          | ███████████████████ 3.62} \\
\texttt{\footnotesize references to female characters or individuals, highlighting | ███████████████████ 3.58} \\
\texttt{\footnotesize references to refugees and related humanitarian crises       | ███████████████████ 3.55} \\
\texttt{\footnotesize instances of the word "seek" and its variations, indicating  | █████████████████ 3.30} \\
\texttt{\footnotesize references to the popularity and significance of sports, par | █████████████████ 3.17} \\
\texttt{\footnotesize references to immigration and asylum issues                  | ████████████████ 3.06} \\
\bottomrule
\end{tabularx}
\caption{Example query from the development set and the top-10 features (with weights) from SPLARE’s latent Bag-of-Concepts representation. Feature explanations are sourced from Neuronpedia~\citep{neuronpedia}.}
\label{tab:table-example}
\end{table*}

%% file: tables/comparison.tex
\begin{table*}[t]
\centering
\begin{tabular}{lc}
\toprule
& nDCG@20 (dev) \\
\midrule
\texttt{\bf Retrieval} \\
MILCO & 0.395 \\
PLAID-X & 0.396 \\
Qwen3-8B-Embed & 0.423 \\
Fusion & 0.468 \\
RUN1 (SPLARE-7B) &  {\bf 0.498}\\
RUN2 (Fusion) &  {\bf 0.517} \\
\midrule 
\texttt{\bf Reranking (pointwise)} \\
Qwen3-Reranker-4B (on Fusion) & 0.487 \\
RUN3 (Qwen3-Reranker-4B on RUN1) &  {\bf 0.545} \\
RUN4 (Qwen3-Reranker-4B on RUN2) &  {\bf 0.548} \\
RUN5 (Fusion RUN2 \& RUN4) &  {\bf 0.559} \\
\bottomrule
\end{tabular}
\caption{Comparison of our five runs to baseline systems - numbers taken from~\citet{lawrie2025neuclirbenchmodernevaluationcollection} - nDCG@20 (dev).}
\label{tab:table-comparison}
\end{table*}

%% file: sections/experiments.tex
\section{Experiments}\label{sec:experiments}

In the following, we describe the five system configurations considered in our study. Configurations marked with $\clubsuit$ correspond to the runs officially submitted, selected based on their performance on the development set.

\subsection{RUN1 - SPLARE}\label{sec:run1}

This configuration corresponds to our baseline retrieval run based on SPLARE-7B~\citep{formal2026learning}. Table~\ref{tab:table-splare} compares the performance of the model without Top-K pooling applied to query and document representations against a pooled variant (Top-K = (40, 400) for queries and documents, respectively—which corresponds to the default evaluation setting in~\citet{formal2026learning}). To maximize effectiveness, we adopt the non-pooled configuration, acknowledging that the pooled variant remains highly competitive (see Table~\ref{tab:table-comparison} for the performance of competitive approaches)—while being much more efficient.

\input{tables/table-splare}

\subsection{RUN2 - SPLARE \& SPLADE fusion}\label{sec:2.1}

We aim to combine SPLARE with a SPLADE model to build an effective first-stage LSR system. We first evaluate SPLADE-8B\footnote{Multilingual SPLADE baseline discussed in~\citet{formal2026learning}.} under the same default configuration (Table~\ref{tab:table-splade}). Interestingly, in the absence of Top-K pooling, SPLADE achieves higher effectiveness than SPLARE, albeit with substantially longer query and document representations. However, under a fixed Top-K pooling configuration, SPLARE consistently outperforms SPLADE, as shown in~\citet{formal2026learning}. Nevertheless, in the context of this challenge, we prioritize maximizing effectiveness and therefore rely on full representations when performing fusion.  
\input{tables/table-splade}
More precisely, we perform a straightforward fusion of SPLARE and SPLADE runs, using Reciprocal Rank Fusion (RRF), implemented with the ranx library~\citep{ranx}, and set the fusion parameter to $k=10$. The resulting system constitutes our second run.

\input{tables/fusion}

\subsection{RUN3 - SPLARE + reranking}

We rerank RUN1 using Qwen3-Reranker-4B~\citep{qwen3embedding}\footnote{\texttt{Qwen/Qwen3-Reranker-4B}.}. Based on a simple hyperparameter sweep (Table~\ref{tab:table-splare-rerank}), we select the top 100 candidates from the initial retrieval stage for reranking. This choice is consistent with prior findings in the reranking literature, which suggest that increasing the number of reranked documents does not necessarily yield additional effectiveness gains.

\input{tables/table-splare-rerank}

\subsection{RUN4 - RUN2 + reranking}

We also rerank RUN2 with Qwen3-Reranker-4B. Similarly to RUN3, we end up reranking 100 candidates (see Table~\ref{tab:table-run2-rerank}).

\input{tables/run2-rerank}

\subsection{RUN5 - RUN2 \& RUN4 fusion}

Finally, as a final run, we recombine using Reciprocal Rank Fusion with $k = 50$ our fused first stage (RUN2) and its reranked list (RUN4). This constitutes our strongest system. 

\input{tables/fusion-final}

%% file: tables/table-splare.tex
\begin{table}[H]
\centering
\begin{tabular}{lc}
\toprule
& nDCG@20 (dev) \\
\midrule
SPLARE-7B | $\text{Top-K}=(40,400)$ & 0.471 \\
SPLARE-7B | no-pooling$^\clubsuit$ & 0.498 \\
\bottomrule
\end{tabular}
\caption{Remark: Average query (resp. document) length $\ell_0(q)=133$ (resp. $\ell_0(d)=457$) for the non-pooled version—compared to $\ell_0(q)=40$, $\ell_0(d)=400$ for the pooled one.}
\label{tab:table-splare}
\end{table}

%% file: tables/table-splade.tex
\begin{table}[H]
\centering
\begin{tabular}{lc}
\toprule
& nDCG@20 (dev) \\
\midrule
SPLADE-8B | $\text{Top-K}=(40,400)$ & 0.464 \\
SPLADE-8B | \text{no-pooling} & 0.512 \\
\bottomrule
\end{tabular}
\caption{Remark: $\ell_0(q)=240$, $\ell_0(d)=807$ for the non-pooled version. At a fixed pooling threshold, SPLADE is less effective than SPLARE.}
\label{tab:table-splade}
\end{table}

%% file: tables/fusion.tex
\begin{table}[H]
\centering
\begin{tabular}{lc}
\toprule
& nDCG@20 (dev) \\
\midrule
SPLARE-7B$^\clubsuit$  & 0.498 \\
SPLADE-8B & 0.512 \\
Fusion 1 (RRF)$^\clubsuit$ & 0.517 \\
\bottomrule
\end{tabular}
\caption{Fusion with RRF - $k=10$. SPLARE and SPLADE are considered without Top-K pooling.}
\label{tab:table-fusion}
\end{table}

%% file: tables/table-splare-rerank.tex
\begin{table}[H]
\centering
\begin{tabular}{lc}
\toprule
  & nDCG@20 (dev) \\
\midrule
SPLARE-7B$^\clubsuit$  & 0.498 \\
\midrule
$k$ & \\
 10 & 0.505  \\
 20 & 0.512  \\
 50 & 0.545  \\
 100$^\clubsuit$ & 0.545  \\
 200 & 0.546   \\
 500 & 0.534  \\
 1000 & 0.536 \\
\bottomrule
\end{tabular}
\caption{Reranking of SPLARE-7B (RUN1) with Qwen3-Reranker-4B - nDCG@20 (dev).}
\label{tab:table-splare-rerank}
\end{table}

%% file: tables/run2-rerank.tex
\begin{table}[H]
\centering
\begin{tabular}{lc}
\toprule
& nDCG@20 (dev) \\
\midrule
Fusion 1 (RRF)$^\clubsuit$ & 0.517 \\
\midrule 
$k$ & \\
10 & 0.522 \\
20 &  0.527 \\
50 &  0.546 \\
100$^\clubsuit$ &  0.548 \\
200 &  0.546 \\
500 & 0.541 \\
1000 &  0.540 \\
\bottomrule
\end{tabular}
\caption{Reranking of our first stage fusion (RUN2) with Qwen3-Reranker-4B - nDCG@20 (dev).}
\label{tab:table-run2-rerank}
\end{table}

%% file: tables/fusion-final.tex
\begin{table}[H]
\centering
\begin{tabular}{lc}
\toprule
& nDCG@20 (dev) \\
\midrule
Fusion 1 (RRF)$^\clubsuit$ & 0.517 \\
Fusion 1 (RRF) $\Rightarrow$ Qwen3 Reranker 4B$^\clubsuit$ & 0.548 \\
Fusion 2 (RRF)$^\clubsuit$ & 0.559 \\
\bottomrule
\end{tabular}
\caption{Fusion with RRF (RUN2 and RUN4) - $k=50$.}
\label{tab:table-fusion-final}
\end{table}

%% file: sections/conclusion.tex
\section{Conclusion}

In this work, we described our approach to the WSDM Cup 2026 shared task on multilingual document retrieval. Our experiments demonstrate that SPLARE, a recently proposed learned sparse retrieval model, achieves performance comparable to leading dense embedding models such as Qwen3-8B-Embed. We further show that integrating score fusion and reranking strategies can significantly enhance retrieval effectiveness, offering promising directions for future research in LSR for multilingual information retrieval.

%% file: tables/examples-2.tex
\begin{table*}[ht]
\centering
\begin{tabularx}{0.85\textwidth}{l}
\toprule
{\footnotesize $\mathcal{Q} \rightarrow$ Corruption during construction of Vostochny Cosmodrome I am looking for reports on officials convicted of} \\ 
{\footnotesize corruption in the construction of the Russian Space launch Complex at Vostochny in the Far East.} \\
\\
\texttt{\footnotesize keywords related to the construction industry                | ████████████████████ 4.12} \\
\texttt{\footnotesize instances and discussions surrounding corruption in various  | ████████████████████ 4.12} \\
\texttt{\footnotesize references to political activism related to rocket launches  | █████████████████ 3.69} \\
\texttt{\footnotesize terms related to complexity in various contexts              | █████████████████ 3.66} \\
\texttt{\footnotesize terms related to criminal convictions and legal proceedings  | █████████████████ 3.64} \\
\texttt{\footnotesize instances of the term "report" and its variations            | ████████████████ 3.41} \\
\texttt{\footnotesize references to government officials and authority figures     | ████████████████ 3.41} \\
\texttt{\footnotesize phrases indicating economic decline and historical context   | ████████████████ 3.36} \\
\texttt{\footnotesize words and phrases related to various forms of construction   | ████████████████ 3.34} \\
\texttt{\footnotesize references to geographical locations, particularly in the ea | ████████████████ 3.33} \\
\midrule
{\footnotesize $\mathcal{Q} \rightarrow$ Danger of Betelgeuse going supernova I am looking for reports on whether} \\ 
{\footnotesize the star Betelgeuse going supernova will present a threat to the Earth.} \\ 
\\
\texttt{\footnotesize mentions of perceived threats in various contexts            | ████████████████████ 4.03} \\
\texttt{\footnotesize occurrences of the name "Betty" and variations of the term " | ███████████████████ 3.88} \\
\texttt{\footnotesize phrases and terms related to danger or hazardous situations  | ██████████████████ 3.67} \\
\texttt{\footnotesize references to Earth and related concepts                     | █████████████████ 3.59} \\
\texttt{\footnotesize mentions of the comedian Rodney Dangerfield                  | █████████████████ 3.47} \\
\texttt{\footnotesize phrases related to movement or transition                    | █████████████████ 3.44} \\
\texttt{\footnotesize instances of the word "present" and its variations           | ████████████████ 3.39} \\
\texttt{\footnotesize references to chemical and physical processes involved in st | ████████████████ 3.36} \\
\texttt{\footnotesize terms related to astronomical events and cosmic phenomena    | ████████████████ 3.34} \\
\texttt{\footnotesize instances of the term "report" and its variations            | ████████████████ 3.33}\\
\midrule
{\footnotesize $\mathcal{Q} \rightarrow$ Bicycles on trains I am looking for information on taking bicycles on local or long-distance trains.} \\ 
\\
\texttt{\footnotesize references to cycling activities and related communities     | ████████████████████ 3.88} \\
\texttt{\footnotesize references to bicycles and cycling-related activities        | ██████████████████ 3.56} \\
\texttt{\footnotesize references to various types of trains and railway systems    | █████████████████ 3.48} \\
\texttt{\footnotesize instances of the word "training" and its variations, indicat | █████████████████ 3.42} \\
\texttt{\footnotesize references to local entities and events in various contexts  | █████████████████ 3.34} \\
\texttt{\footnotesize references to railroads and related infrastructure           | ████████████████ 3.25} \\
\texttt{\footnotesize references to trains or train-related concepts               | ███████████████ 3.03} \\
\texttt{\footnotesize terms related to bicycles and their use in context           | ███████████████ 3.03} \\
\texttt{\footnotesize phrases and terms related to distance and measurement        | ███████████████ 3.02} \\
\texttt{\footnotesize references to luggage and baggage handling processes         | ██████████████ 2.89} \\
\bottomrule
\end{tabularx}
\caption{Example queries from the development set and the top-10 features (with weights) from SPLARE’s latent Bag-of-Concepts representation. Feature explanations are sourced from Neuronpedia.~\citep{neuronpedia}.}
\label{tab:table-example-2}
\end{table*}